\documentclass[aip,jcp,reprint,amsmath,amssymb,floatfix,citeautoscript]{revtex4-2}
\usepackage{cancel}
\usepackage{amsmath}
\usepackage{physics}
\usepackage{footmisc}
\usepackage{graphicx}
\usepackage{amssymb}
\usepackage{sidecap}
\usepackage{amsthm}
\usepackage{bm}
\usepackage{dcolumn}% Align table columns on decimal point
\usepackage{braket}% Enables braket notation
\usepackage{longtable}
\usepackage{multirow}
\usepackage{booktabs}
\usepackage{ragged2e}
\usepackage{txfonts}
\usepackage[usenames,dvipsnames]{xcolor}
\definecolor{coquelicot}{rgb}{1.0, 0.22, 0.0}
\usepackage[colorlinks=true,allcolors=blue]{hyperref}

\newcommand{\Mact}{M_{\mathrm{act}}}
\newcommand{\vk}{{\bm{k}}}

\begin{document}

\title{Dynamical correlation energy of metals in large basis sets from downfolding and composite approaches}

\author{James M. Callahan}
\affiliation{Department of Chemistry, Columbia University, New York, New York 10027 USA}
\author{Malte F. Lange}
\affiliation{Department of Chemistry, Columbia University, New York, New York 10027 USA}
\author{Timothy C. Berkelbach}
\affiliation{Department of Chemistry, Columbia University, New York, New York 10027 USA}
\affiliation{Center for Computational Quantum Physics, Flatiron Institute, New York, New York 10010 USA}
\email{tim.berkelbach@gmail.com}

\date{\today}

\begin{abstract}

Coupled-cluster theory with single and double excitations (CCSD) is a promising
\textit{ab initio} method for the electronic structure of three-dimensional metals, for which
second-order perturbation theory (MP2) diverges in the thermodynamic limit.
However, due to the high cost and poor convergence of CCSD with respect to basis size,
applying CCSD to periodic systems often leads to large basis set errors.
In a common ``composite'' method, MP2 is used to recover the missing dynamical correlation energy
through a focal-point correction,
but the inadequacy of MP2 for metals raises questions about this approach.
Here we describe how high-energy excitations treated by MP2 can be ``downfolded'' into a low-energy
active space to be treated by CCSD.
Comparing how the composite and downfolding approaches perform for the uniform electron
gas, we find that the latter converges more quickly with respect to the basis set size.
Nonetheless, the composite approach is surprisingly accurate because
it removes the problematic MP2 treatment of double excitations near the Fermi surface.
Using the method to estimate the CCSD correlation energy in the combined complete basis set and thermodynamic
limits, we find CCSD recovers over 90\% of the exact correlation energy at $r_s=4$.
We also test the composite and downfolding approaches with the random-phase approximation used in place of MP2,
yielding a method that is more effective but more expensive. 

\end{abstract}

\maketitle

\section*{Introduction}
Ground-state electronic properties of metallic solids have traditionally been
computed using density functional theory~(DFT)~\cite{Kohn1965,Kresse1996,Burke2012,Jones2015},
which is partially justified by the fact that many popular functionals are parameterized by 
numerically exact results on the uniform electron gas (UEG)~\cite{Ceperley1980,Huang2008,Giuliani2008,Loos2016}.
In recent years,
interest has grown around the application of \textit{ab initio},
wavefunction-based electronic structure techniques for condensed-phase
systems~\cite{Pisani2008,Gruneis2010,Shepherd2012a,Booth2013,Yang2014,McClain2017,Gruber2018,Brandenburg2019},
since they do not suffer
from uncontrolled errors inherent to the DFT exchange-correlation
functional~\cite{Huang2008,Tsatsoulis2017,Usvyat2018}.
Promising methods in this direction include the random-phase approximation (RPA)~\cite{Harl2009,Gruneis2009,Brandenburg2019}
and coupled-cluster theory~\cite{Shavitt2009,Bartlett2012,Zhang2019,Gruneis2020,Stoll2009,Hummel2018}.
Importantly,
both of these methods
preclude the well-known divergences of finite-order perturbation theories,
such as second-order M\o{}llet-Plesset perturbation theory (MP2),
via an infinite-order resummation~\cite{Gell-Mann1957,Bohm1957,Freeman1977,Bartlett2007,Shepherd2013}.

Although coupled-cluster theory has been successfully applied to an increasing number of
atomistic semiconductors and insulators~\cite{Liao2016,McClain2017,Gruber2018,Gruber2018a,Dittmer2019,Gao2020,Wang2020,Pulkin2020},
its applicability for metals has been primarily focused around the UEG,
also known as jellium~\cite{Freeman1977,Freeman1978,Freeman1983,Shepherd2016,McClain2016,Spencer2016,Neufeld2017,Lewis2019,White2020}.
Despite their reasonable accuracy,
these calculations have demonstrated the typical
slow convergence of the correlation energy as a function of the number of virtual (unoccupied) orbitals
included~\cite{Shepherd2012a,Shepherd2012b,Hattig2012,Shepherd2016}.
This slow convergence is especially problematic because of the high cost of
coupled-cluster calculations with large basis sets.
For example,
coupled-cluster theory with single and double
excitations (CCSD) has a computational cost
that scales as $O(N^2M^4)$,
where $N$ and $M$ are the number of electrons and basis functions respectively.
To date,
results near the complete basis set (CBS) limit have been primarily computed via
the extrapolation of results obtained with a finite,
increasing number of basis functions~\cite{Shepherd2016,White2020,Gruneis2020},
although explicitly correlated~\cite{Gruneis2013} and
transcorrelated~\cite{Luo2018,Liao2021arxiv} methods provide promising alternative approaches.

Composite methods
(sometimes called focal point methods)
are a simple,
alternative class of approaches for
recovering dynamical correlation within large basis sets~\cite{Fiedler2017,Kumar2017,Warden2020}.
A common composite scheme combines the results of high-level and low-level theories using three calculations.
For example,
using CCSD as the high-level theory and MP2 as a low-level theory,
the CCSD correlation energy in a large basis is approximated as
\begin{equation}
\label{eq:comp}
E_\mathrm{CCSD}(M) \approx E_\mathrm{CCSD}(\Mact) 
    + E_\mathrm{MP2}(M) - E_\mathrm{MP2}(\Mact)
\end{equation}
where $\Mact < M$ is the number of ``active'' basis functions.
Refs.~\onlinecite{Irmler2019,Irmler2019a} provide a similar but more sophisticated CCSD/MP2 composite method,
based on an analysis of the basis set convergence of various diagrammatic contributions 
to the correlation energy.
While such CCSD/MP2 composite approaches have been applied successfully to a number of semiconductors 
and insulators~\cite{Irmler2019,Lange2020,Wang2020,Lau2021},
their applicability to metals is questionable because of the failures of MP2 theory.
One goal of this work is to test the composite CCSD/MP2 approach for metals.

A more theoretically satisfying approach would be to perform a single
calculation where low-energy excitations near the Fermi surface are treated with CCSD
and are coupled to high-energy excitations treated with MP2.
This particular approach, which is similar to tailored CC~\cite{Kinoshita2005}
and the broader class of active-space CC
methods~\cite{Piecuch1999,Kowalski2001,Piecuch2010,Shen2012,Dutta2017},
has variously been called CC/PT~\cite{Nooijen1999},
CCSD-MP2~\cite{Bochevarov2005},
and multilevel CC~\cite{Myhre2013,Myhre2014}.
Two of us (M.F.L.~and T.C.B.)
recently tested this method for a few simple atomistic semiconductors and insulators~\cite{Lange2020},
and here we aim to assess its performance for metals,
where the differences between CCSD and MP2 are more striking.
Since the effects of the frozen high-energy MP2 amplitudes are folded down onto the low-energy
CCSD amplitudes (see below),
we refer to this method as a
``downfolding'' approach.
In principle,
this downfolding CCSD/MP2 method
should provide a distinct advantage over the conceptually simpler
composite approach, 
as downfolding does not include the MP2 treatment of low-energy excitations
that are responsible for divergence in the thermodynamic limit (TDL).
After providing theoretical details of these two methods, we compare their
performance for the UEG at a fixed number of electrons and in the TDL. 
Before concluding, we also examine the straightforward use of the RPA in place of
MP2.

\section*{Theory}
Here we briefly review the theory underlying the downfolding and composite approaches.
The $N$ occupied spin-orbitals are indexed by $i,j,k,l$;
the $\left(M-N\right)$ virtual orbitals by $a,b,c,d$;
and the $M$ general orbitals by $p,q,r,s$.
The MP2 and coupled-cluster with double excitations (CCD) correlation energies are given by
\begin{equation}
E_\mathrm{c} = \frac{1}{4} \sum_{ijab} t_{ij}^{ab} \langle ij \vert \vert ab \rangle
\label{eq:ecorr}
\end{equation}
where $t_{ij}^{ab}$ are amplitudes of the double excitation operator 
$T_2 = \frac{1}{4} \sum_{ijab} t_{ij}^{ab} a_a^{\dagger} a_b^{\dagger} a_j^{} a_i^{}$
and $\langle pq||rs\rangle$ are antisymmetrized two-electron repulsion integrals;
contributions from single excitations vanish because the UEG has no capacity for orbital
relaxation by symmetry.
At lowest order in perturbation theory,
\begin{equation}
t_{ij}^{ab} 
    = \frac{\langle ab \vert \vert ij \rangle}{\varepsilon_i + \varepsilon_j - \varepsilon_a - \varepsilon_b},
\label{eq:MP2}
\end{equation}
and Eq.~(\ref{eq:ecorr}) gives the MP2 correlation energy.
The high density of states at the Fermi surface and the long-ranged nature of the Coulomb potential
are together responsible for the divergence of MP2 in the TDL.
By contrast, the CCD amplitudes solve a system of nonlinear equations
\begin{equation}
0 = \langle \Phi_{ij}^{ab} \vert e^{-T_2} H e^{T_2} \vert \Phi \rangle,
\label{eq:CCD}
\end{equation}
where $H$ is the electronic Hamiltonian.
A standard approach for reaching the CBS limit is to
perform a series of calculations with increasing $M$ and use a $M^{-1}$ extrapolation. 

In both the composite and downfolding approaches,
we partition the orbitals into a set of $\Mact$ active orbitals, composed of 
all occupied orbitals and the low-energy virtual orbitals, and a set of
$\left(M-\Mact\right)$ frozen (inactive) orbitals, composed of the high-energy virtual orbitals.
In principle, occupied orbitals can also be partitioned,
but typically they do not significantly contribute
to the computational cost.
In the composite CCD/MP2 approach, the correlation energy is calculated according
to Eq.~(\ref{eq:comp}).
Importantly for metals, the low-energy active space double excitations are 
treated by CCD and not by MP2, so we expect the method to be well-behaved in the
thermodynamic limit.

In the downfolding CCD/MP2 approach, the double excitation operator $T_2$ is
partitioned into internal excitations fully contained within the active space
and external excitations that involve at least one frozen orbital,
$T_2 = T_2^{\mathrm{(int)}} + T_2^{\mathrm{(ext)}}$.
Fixing the $T_2^{\mathrm{(ext)}}$ amplitudes to their MP2 values via Eq.~(\ref{eq:MP2}),
the downfolding method involves first solving Eqs.~(\ref{eq:CCD}) for
\textit{only} the internal amplitudes and then evaluating the correlation energy
expression Eq.~(\ref{eq:ecorr}) using \textit{both} the internal and external
amplitudes.
Compared to the $O(N^2 M^4)$ cost of full CCD,
the composite approach has $O(N^2 \Mact^4)+O(N^2 M^2)$ cost and
the downfolding approach has $O(N^2 \Mact^2 M^2)$ cost,
which can provide significant savings,
depending on the practical value of the ratio $\Mact/M$.

Let us now provide more insight into the ``downfolding'' perspective.
Note that,
because the internal and external excitation operators commute,
the defining energy and amplitude equations of the downfolding 
approach can also be written
\begin{subequations}
\label{eq:downfolding}
\begin{align}
\begin{split}
E_\mathrm{c} &= \langle \Phi | e^{-T_2^{\mathrm{(int)}}} (\bar{H}-E_\mathrm{HF}) e^{T_2^{\mathrm{(int)}}} | \Phi \rangle \\
    &= E_\mathrm{MP2}^\mathrm{(ext)} 
    + \frac{1}{4}\sum_{ijab}^{\mathrm{active}} t_{ij}^{ab} \langle ij||ab\rangle 
\end{split} \\
0 &= \langle \Phi_{ij}^{ab} \vert e^{-T_2^{\mathrm{(int)}}} \bar{H} e^{T_2^{\mathrm{(int)}}} \vert \Phi \rangle
\quad (i,j,a,b)\ \mathrm{active}
\label{eq:downfolding_amp}
\end{align}
\end{subequations}
where $\bar{H} = e^{-T_2^{\mathrm{(ext)}}} H e^{T_2^{\mathrm{(ext)}}}$
(with fixed $T_2^{\mathrm{(ext)}}$ as detailed above) and
$E_\mathrm{MP2}^\mathrm{(ext)} = \frac{1}{4} \sum_{ijab}^{\mathrm{(ext)}} t_{ij}^{ab} \langle ij||ab\rangle$ 
is the MP2 correlation energy due to external excitations.
These resemble ordinary CCD energy and amplitude equations 
\textit{within the active space only},
except that the bare Hamiltonian $H$ is replaced by an 
\textit{effective} Hamiltonian $\bar{H}$ that is similarity-transformed by the 
external excitation amplitudes.
The effective Hamiltonian within the active space can be expressed as
\begin{equation}
\label{eq:heff}
\begin{split}
    \bar{H} - E_\mathrm{HF} &= E_\mathrm{MP2}^\mathrm{(ext)}
    + \sum_{pq}^{\textrm{active}} F_{pq} \lbrace a_p^\dagger a_q \rbrace \\
&\hspace{1em} + \frac{1}{4} \sum_{pqrs}^{\textrm{active}} W_{pqrs} 
    \lbrace a_p^\dagger a_q^\dagger a_s a_r \rbrace + \dots,
\end{split}
\end{equation}
where $\lbrace\dots\rbrace$ indicates normal ordering of the operators.
This effective Hamiltonian can be seen to contain effective one- and two-body
interactions that are frequency independent~\cite{Shavitt2009},
in contrast to other downfolding approaches like the constrained random-phase
approximation~\cite{Aryasetiawan2004,Miyake2008}.
For example, the all-occupied two-body interaction becomes
\begin{equation}
W_{ijkl} = \langle ij||kl\rangle 
    + \frac{1}{2} \sum_{ab}{\vphantom{\sum}}^\prime 
        \frac{\langle ij||ab\rangle \langle ab||kl\rangle}
            {\varepsilon_k + \varepsilon_l - \varepsilon_a - \varepsilon_b},
\end{equation}
where the primed summation indicates that one or both of $a,b$ are inactive virtual orbitals.
The frequency independence can be understood because our observable is the total energy
rather than a spectral function.
To summarize, an approach that solves the internal CCD amplitude equations in
the presence of frozen external amplitudes is equivalent to a
CCD calculation in an active space of orbitals using an effective (downfolded)
Hamiltonian that is similarity-transformed by the external excitation operator.

It is straightforward to show that the composite approach,
normally understood
as a three-step procedure as shown in Eq.~(\ref{eq:comp}),
is equivalent to Eqs.~(\ref{eq:downfolding})
but where the effective Hamiltonian
$\bar{H}$ is replaced by the bare Hamiltonian $H$ in the amplitude
equations~(\ref{eq:downfolding_amp}).
From this perspective,
the performance differences between the downfolding and composite approaches
are attributable to the screening of the integrals in the effective
Hamiltonian when determining the internal amplitudes.
Nevertheless, we reiterate that the composite CCD/MP2 approach is expected to perform well because it 
replaces the problematic MP2 treatment of low-energy, internal double excitations with a
well-behaved CCD treatment.

\begin{figure}
    \centering
    \includegraphics{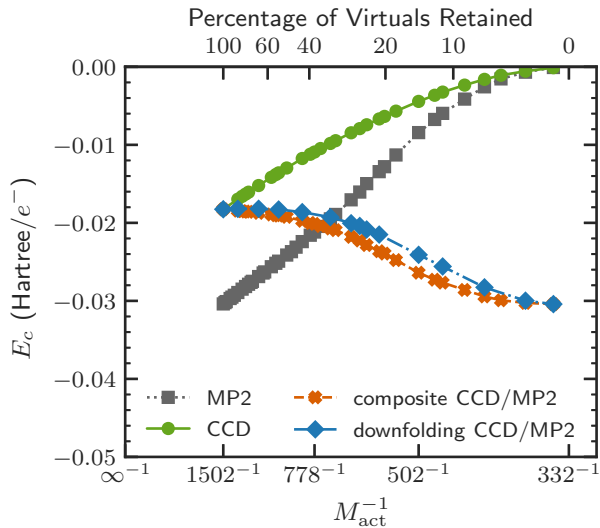}
    \caption{Basis set convergence of the correlation energy of the $r_s=4$ UEG with $N=332$~electrons
    for MP2 (squares), CCD (circles), composite CCD/MP2 (crosses), and downfolding CCD/MP2 (diamonds).
    The top axis shows the percentage of virtual orbitals that are active for the composite and downfolding methods,
    compared to the ``target'' calculation with $M=1502$.
    Both the composite and downfolding methods interpolate between the ``target'' MP2 calculation (leftmost square)
    and the ``target'' CCD calculation (leftmost circle).
    }
    \label{fig:A}
\end{figure}

\section*{Results and discussion}
We study the UEG as the simplest model of metals.
A brief review of the UEG model in finite cells with finite plane-wave basis sets is given in the Appendix 
and we refer the reader to the literature for more
details~\cite{Shepherd2013,Shepherd2014,Shepherd2016}.
To illustrate the performance of the composite and downfolding methods,
we focus on the Wigner-Seitz radius $r_s = 4$
(corresponding to the approximate valence electron density of metallic sodium),
where CCD has been found to recover about 85\% of the correlation energy~\cite{Shepherd2016,Neufeld2017}.
We use a twisted boundary condition by performing calculations at the Baldereschi point~\cite{Baldereschi1973},
which has been shown to provide smoother convergence to the TDL~\cite{Drummond2008,Mihm2019}.

In Fig.~\ref{fig:A}, we show basis set convergence of the correlation energy
for a finite UEG with $N = 332$~electrons.
The uncorrected MP2 and CCD correlation energies exhibit their typical slow
convergence with increasing basis set size
and show asymptotic behavior where the basis set error decays as $M^{-1}$.
Extrapolation to the CBS limit yields $E_\mathrm{c}/N = -0.0401~E_\mathrm{h}$ for MP2
and $E_\mathrm{c}/N = -0.0262~E_\mathrm{h}$ for CCD.
At the largest finite basis shown, $M = 1502$,
the results exhibit a significant basis set error of about $0.01~E_\mathrm{h}$ for both methods,
highlighting the challenge of recovering dynamical correlation in metals with large basis sets.
Importantly, we emphasize that the MP2 correlation energy
does not diverge \textit{for any finite system} but only upon extrapolation to the TDL (see below).

Recall that the CCD/MP2 composite and downfolding approaches involve both a ``target''
number of orbitals $M$
and an active number of orbitals $\Mact$.
In Fig.~\ref{fig:A},
we show results obtained for $M = 1502$ as $\Mact$ is varied.
By construction,
both methods yield the target MP2 correlation energy when there are no active 
virtual orbitals ($\Mact = N$)
and the target CCD correlation energy when all orbitals are active ($\Mact=M$). 
We observe that both methods converge
smoothly to the target CCD result
and that the downfolding approach exhibits a faster convergence,
due to its coupling between the internal and external excitation spaces.
We also see similar behavior for other numbers of electrons and densities (not shown),
indicating that neither finite-size effects nor the specific metallic
density changes the overall picture.

\begin{figure}
    \centering
    \includegraphics{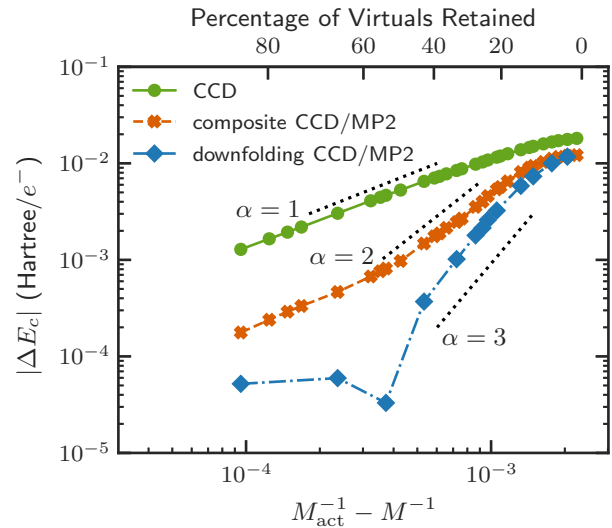}
    \caption{Absolute error in the correlation energy for the data in Figure~\ref{fig:A},
    shown on a logarithmic scale,
    relative to the ``target'' CCD result with $M = 1502$.
    The top axis and symbols have the same meaning as Figure~\ref{fig:A}.
    Dotted black lines are shown as a guide for various power law exponents $\alpha$ as discussed in the text.
    }
    \label{fig:B}
\end{figure}

To better quantify the rate of convergence,
in Fig.~\ref{fig:B} we plot the
absolute deviation of the correlation energy from the ``target'' CCD result obtained with $M = 1502$.
The error is plotted as a function of the difference between the inverse number of active
orbitals and the inverse number of total orbitals
and analyzed in terms of the
power law $|\Delta E_\mathrm{c}| \propto \left[\Mact^{-1}-M^{-1}\right]^\alpha$.
We compare the convergence of traditional CCD,
the composite approach,
and the downfolding approach.
For plain CCD, we see linear convergence of the correlation energy,
with $\alpha \approx 1$,
over a large range of basis set sizes.
The composite method exhibits an early,
rapid convergence reaching a maximum scaling of around $\alpha \approx 2$
before slowing to the same $\alpha \approx 1$ convergence
as $\Mact$ approaches $M$.
The rapid convergence for small $\Mact$ is responsible for
absolute errors that are about one order of magnitude better than those obtained by simple
truncation.
In fact,
the plain CCD result does not obtain m$E_\mathrm{h}$ accuracy until essentially
all orbitals are correlated,
whereas the composite result achieves this accuracy when only
50\% of the virtual orbitals are correlated in the expensive CCD calculation;
this results in
a speedup of a factor of 16 compared to the full CCD calculation.
Finally, the downfolding result exhibits rapid but non-monotonic convergence,
making it difficult to extract a power law.
Before slightly overshooting the ``target'',
the power law exponent reaches $\alpha \approx 3$ or better,
a significant improvement over the composite CCD/MP2 and standard CCD approaches.
This fast rate of convergence
provides m$E_\mathrm{h}$ accuracy when about one third of the virtual orbitals are correlated,
giving a speedup of a factor of 9.

\begin{figure}[t]
    \centering
    \includegraphics{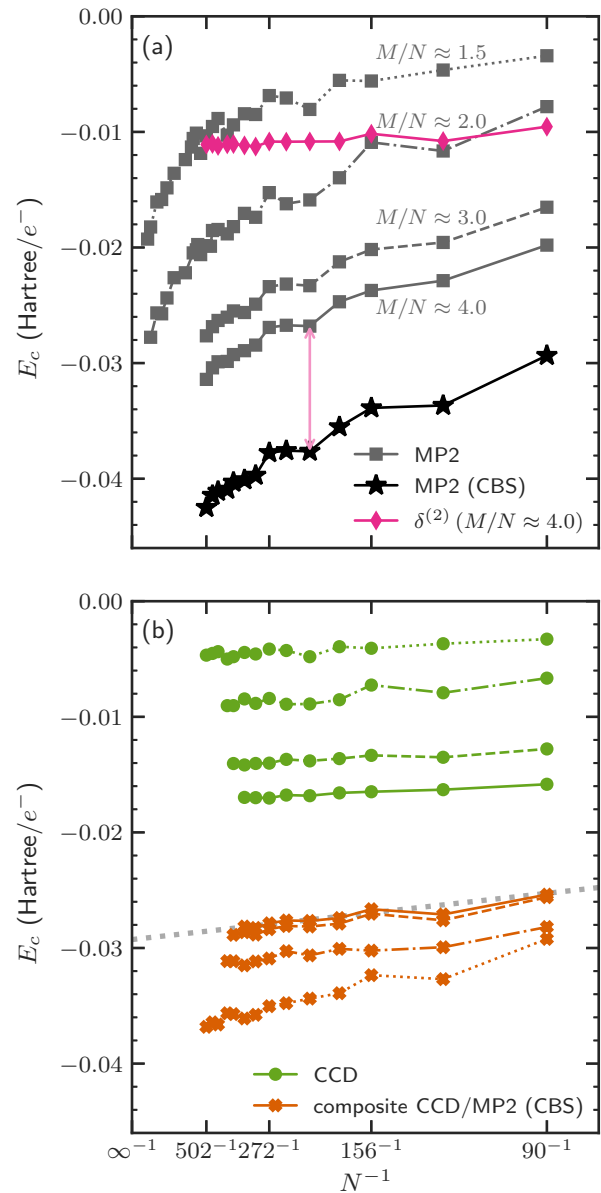}
    \caption{Thermodynamic limit convergence of the correlation energy of the $r_s=4$ UEG for 
    systems with $N=90$--2392~electrons.
    (a) MP2 results for basis set sizes indicated (squares) and in the extrapolated CBS limit (stars).
    Thin diamonds show the CBS MP2 correction, $\delta^{(2)}\left(M/N \approx 4.0 \right)$,
    indicated by the double-headed arrow at $N=210$.
    (b) CCD (circles) and composite CCD/MP2 (crosses) results at the same (active) basis set sizes.
    For composite CCD/MP2, we applied the CBS MP2 correction $\delta^{(2)}\left(M\right)$.
    The grey line shows our TDL extrapolation using the largest five systems for $M/N \approx 4.0$.
    }
    \label{fig:D}
\end{figure}

The good performance of the composite approach indicates that MP2,
while an inapplicable theory for three-dimensional metals,
is safe to use for basis set corrections.
As discussed above, the reason for this applicability can be understood
by considering the MP2 correction that is applied
in Eq.~(\ref{eq:comp}).
This correction is a difference between two MP2 correlation energies,
\textit{both} of which correlate orbitals near the Fermi surface.
These two MP2 energies are separately divergent in the TDL,
but their difference is not;
moreover, this difference is precisely $E_\mathrm{MP2}^{(\mathrm{ext})}$
defined previously.

The reliable scaling of the MP2 basis set error at large $M$
suggests that the MP2 CBS limit can be obtained by extrapolation
for any given number of electrons.
In contrast,
the asymptotic scaling regime for the CCD correlation energy
cannot always be reached.
Thus,
for any calculation performed with a given $M$,
we propose to add the MP2 correlation energy difference
$\delta^{(2)}\left(M\right) = E_\mathrm{MP2}\left(\infty\right) - E_{\mathrm{MP2}}\left(M\right)$,
where $E_\mathrm{MP2}\left(\infty\right)$
is obtained by $M^{-1}$ extrapolation.
Having obtained an estimate of the CBS limit for a given number of electrons,
the finite-size extrapolation to the TDL can be done separately.
We expect this scheme to be not only more reliable than extrapolating CCD on its own
but also less costly since it involves more MP2 and fewer CCD calculations.

In Fig.~\ref{fig:D}(a),
we plot the MP2 correlation energy
as a function of the inverse number of electrons 
for finite UEG systems containing $N = 90$--2392 electrons.
For each system size, we performed MP2 calculations at different basis set sizes (grey squares),
and the top four grey curves connect systems with a similar ratio of $M/N$. 
We then performed $M^{-1}$ extrapolations at each particle number using data 
from 
$M/N \approx 3.0, 3.5, 4.0, 4.5$
to obtain the CBS limit at each system size (black stars).
On approach to the TDL, 
the MP2 correlation energy diverges, as seen most easily in our largest calculations for the smaller
values of $M/N$.
Despite the divergence of its components,
the MP2 CBS correction $\delta^{(2)}$,
plotted explicitly in Fig.~\ref{fig:D}(a) for $M/N \approx 4$ (pink thin diamonds),
does \textit{not} diverge and is thus safe for use in metallic systems.

To confirm that the success of the composite CCD/MP2 method holds on approach to the TDL,
in Fig.~\ref{fig:D}(b) we plot results for CCD (green circles) and composite CCD/MP2 (orange crosses).
For the composite CCD/MP2 result, we applied the MP2 CBS correction $\delta^{(2)}\left(M\right)$,
although we note that a similar CBS correction could also be combined with the downfolding CCD/MP2 approach.
In contrast to MP2, CCD is well-defined in the TDL;
however, at each system size, convergence to the CBS limit is slow,
as shown by the green data sets.
In contrast, we observe that composite CCD/MP2 has much faster convergence to the TDL,
as shown by the orange data sets.
Performing a $N^{-1}$ extrapolation of the CBS-corrected $M/N \approx 4.0$ results 
gives us a CBS and TDL extrapolated correlation energy of $E_\mathrm{c} = -0.0293~E_\mathrm{h}$,
in general agreement with past CCD results\cite{Shepherd2016}.  For comparison,
the exact value~\cite{Vosko1980} is $E_\mathrm{c} = -0.0318~E_\mathrm{h}$, indicating that CCD 
recovers over 90\% of the correlation energy.
The same finite-size extrapolation of the CBS-corrected $M/N \approx 3.0$ data gives a correlation energy
that differs by only $0.7~\mathrm{m}E_\mathrm{h}$,
indicating the excellent convergence of the
composite method.

\begin{figure}
    \centering
    \includegraphics{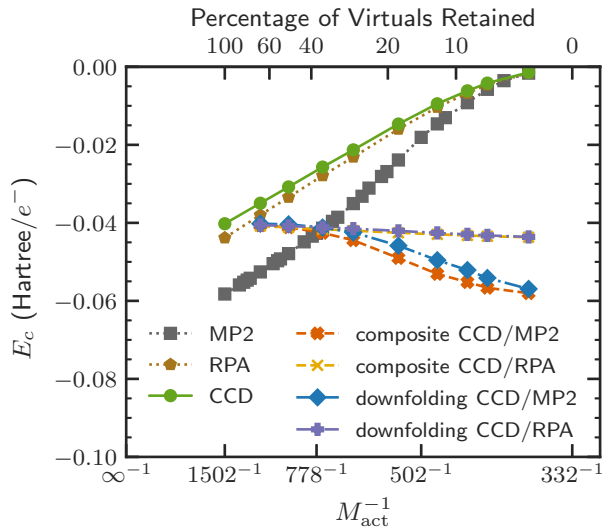}
    \caption{The same as in Fig.~\ref{fig:A}, except for $r_s=1$.
    In addition to the methods shown in Fig.~\ref{fig:A},
    we also include RPA (pentagons), composite CCD/RPA (thin crosses), and downfolding CCD/RPA (pluses).
    }
    \label{fig:E}
\end{figure}

Before concluding,
we recognize that a variety of other low-level theories can be combined with CCD,
in both the composite and downfolding manner.
Of particular interest is the RPA,
which is more appropriate for
metals than MP2 but also more expensive.
The so-called direct RPA is most promising computationally due
to its low cost when two-electron integrals are handled by density fitting.
Here we instead test the full particle-hole RPA,
where the $T_2$ amplitudes maintain proper anti-symmetry
so that the theory is free of self-interaction error.
The RPA amplitudes are the solution of the CCD equations where only
selected terms are retained~\cite{Scuseria2008,Shepherd2014}.
In Fig.~\ref{fig:E},
we show the same results as in Fig.~\ref{fig:A}, except for the electron density corresponding to $r_s=1$
(due to RPA convergence problems at larger values of $r_s$~\cite{Shepherd2014}).
We observe the same overall trends as we did at $r_s=4$ for the MP2 and CCD calculations with
finite basis sets.
However, the RPA results follow the CCD results much more closely,
which also makes composite and downfolding CCD/RPA significantly outperform
the corresponding MP2 methods.
Unsurprisingly, the improved performance of the downfolding approach compared to the composite approach
is even more marginal than for CCD/MP2.

\section*{Conclusions and outlook}
We have described and analyzed two approaches for eliminating basis set error in the CCD correlation energy
of metals, using the simple UEG model.
Our results indicate that these methods allow for aggressive freezing of virtual orbitals or approximation
of external amplitudes,
leading to significant reductions in computational cost.
Although the downfolding CCD/MP2 approach is slightly more accurate,
we find that the simpler composite CCD/MP2 approach is surprisingly effective
because divergent contributions near the Fermi surface do not
contribute to the basis set correction.

In this work, we have addressed post-Hartree-Fock basis set errors in the canonical orbital basis,
but the methods we presented could also be straightforwardly applied in a basis of localized orbitals.  
For example, Refs.~\onlinecite{Schafer2021,Lau2021}
used CCSD/MP2 and CCSD/RPA composite approaches to mitigate basis set errors in quantum embedding
calculations.
Localized orbital basis sets mix orbitals near and far from the Fermi surface,
so it will be interesting to test how the composite and downfolding approaches perform
when using these localized orbitals for metals.

Future work will focus on applying these techniques to atomistic metals using
natural orbitals~\cite{Gruneis2011}, to the excited-state properties
of metals~\cite{McClain2016,Lewis2019}, and to higher-level theories
of correlation~\cite{Neufeld2017}.
For example, we imagine that
a composite CCSDT/CCSD or CCSDT/CCSD(T) approach would provide quantitative accuracy for metals while
precluding the failure~\cite{Shepherd2013} of the otherwise successful treatment of 
perturbative triple excitations.

\begin{acknowledgments}
We thank Verena Neufeld and Sandeep Sharma for comments on
the manuscript.
This work was supported in part by the National Science Foundation Graduate
Research Fellowship under Grant No.~DGE-1644869 (M.F.L.), by the Department of Defense
through the National Defense Science \& Engineering Graduate (NDESG) Fellowship Program (J.M.C.), and
by the National Science Foundation under Grant No.~CHE-1848369 (T.C.B.). 
We acknowledge computing resources from Columbia
University’s Shared Research Computing Facility project, which is supported by
NIH Research Facility Improvement Grant 1G20RR030893-01, and associated funds
from the New York State Empire State Development, Division of Science Technology
and Innovation (NYSTAR) Contract C090171, both awarded April 15, 2010.
The Flatiron Institute is a division of the Simons Foundation.
\end{acknowledgments}

\section*{Appendix: Uniform electron gas} \label{app:ueg}
Working in a single-particle basis of plane waves with momenta $\vk=(2\pi/L)(n_x, n_y, n_z)$
and cell of volume $L^3$ with periodic boundary conditions,
the UEG Hamiltonian is given by
\begin{equation}
\begin{split}
H &= \sum_{\vk\sigma} \frac{k^2}{2} a_{\vk\sigma}^\dagger a_{\vk\sigma} \\
    &+ \frac{1}{2} \sum_{\vk_1\vk_2\vk_3\vk_4}^\prime \sum_{\sigma\sigma^\prime} 
        \langle \vk_1\sigma,\vk_2\sigma^\prime|\vk_3\sigma,\vk_4\sigma^\prime\rangle
        a^\dagger_{\vk_1\sigma} a^\dagger_{\vk_2\sigma^\prime} a_{\vk_4\sigma^\prime} a_{\vk_3\sigma}
\end{split}
\end{equation}
where the primed summation requires $\vk_1 +\vk_2 = \vk_3 + \vk_4$.
The two-electron repulsion
integrals are given by
\begin{equation}
\langle \vk_1\sigma,\vk_2\sigma^\prime|\vk_3\sigma,\vk_4\sigma^\prime\rangle
    = v(\vk_1-\vk_3) \delta_{\vk_1+\vk_2, \vk_3+\vk_4}
\end{equation}
where the Ewald potential is 
\begin{equation}
\begin{aligned}
v(\vk) &=
\begin{cases}
\dfrac{4 \pi}{L^3 k^2} & k \neq 0
\\ \\
v_\mathrm{M} & k = 0,
\end{cases}
\end{aligned}
\label{eq:madelung}
\end{equation}
and $v_\mathrm{M} = 2.837297479/L$ is the Madelung constant of the cell~\cite{Sholl1967}.
The $N$-electron reference determinant has
the lowest-energy $N/2$ plane wave orbitals doubly occupied
and the HF orbital energies are given by
$\varepsilon(\vk) = k^2/2 - v_\mathrm{M}\theta(k_\mathrm{F}-k)$,
where $k_\mathrm{F}$ is the Fermi momentum.
We restrict our calculations to closed-shell configurations,
which allows only certain ``magic numbers'' of electrons and orbitals.
At the Baldereschi point,
the first few magic numbers are $2,8,14,22,34,40,52$.

\section*{Data availability statement}
The data that support the findings of this study are available
from the corresponding author upon reasonable request.

%\bibliography{jamesbib_fixed.bib,preprints.bib}
%

\end{document}